\documentclass[%
nofootinbib,
aps,
prd,
floatfix,
twocolumn,
]{revtex4-2}

\usepackage{dcolumn}
\usepackage[mathscr]{eucal}
\usepackage{tensind}
\tensordelimiter{?}
\usepackage[fleqn]{amsmath}
\usepackage{amssymb}
\usepackage{bm}
\usepackage{ulem}
\usepackage{graphicx}
\usepackage{subcaption}
\usepackage{xcolor}
\usepackage{microtype}
\usepackage{mathtools} 
\usepackage{upgreek}
\usepackage{hyperref}
\usepackage{lipsum}
\usepackage{xifthen} 

\usepackage{general_sty}
\usepackage{mathematical_shorthands}
\usepackage{continuum_mechanics_definitions}

\newcommand{\tdot}{\,.\hspace{-0.98 mm}\raise.6ex\hbox{.}
                   \hspace{-0.98 mm}\raise1.2ex\hbox{.}\,}

\newcommand{\Ghat}{\hat{G}}
\newcommand{\Gcirc}{\mathring{G}}
\newcommand{\covst}[1]{\breve{\nabla}_{#1}}
\newcommand{\covmet}[1]{\hat{\nabla}_{#1}}
\newcommand{\Ct}{\widetilde{C}}

\newcommand{\SQuad}{\mathfrak{S}}
\renewcommand{\wp}{w^{\perp}}
\newcommand{\Thp}{\Theta^{\perp}}
\newcommand{\wtp}{\tilde{w}^{\perp}}
\newcommand{\Gamhat}{\hat{\Gamma}}

\newcommand{\christMom}{[\Gamhat\,\Theta]}
\newcommand{\christMTheta}{{\Gamma_{\Theta}}}
\newcommand{\christmu}{{\Gamma_{M}}}
\newcommand{\christtheta}{{\Gamma_{M}}}

\newcommand{\coupStress}{%
                         \lambda%
                        }
\newcommand{\CoupStress}{%
                         \Lambda%
                        }
\newcommand{\massMom}{\pi}
\newcommand{\MassMom}{\Pi}

\newcommand{\EGR}{EGR}
\newcommand{\ECGR}{EC theory}

\renewcommand{\wording}[1]{{#1}}
\renewcommand{\edit}[1]{{#1}}

\begin{document}

\title{Cosserat elasticity as the weak-field limit of Einstein--Cartan relativity}

\author{Matthew Maitra}
\email{matthew.maitra@erdw.ethz.ch}
\affiliation{%
  Institut f\"{u}r Geophysik,
  ETH Z\"{u}rich,
  Sonneggstrasse 5,
  Z\"{u}rich 8092,
  Switzerland
}
\author{Jeroen Tromp}
\email{jtromp@princeton.edu}
\affiliation{%
 Department of Geosciences and
 Program in Applied and Computational Mathematics,
 Princeton University,
 Princeton, New Jersey 08540,
 USA
}

\date{\today}

\begin{abstract}
The weak-field limit of Einstein--Cartan (EC) relativity is studied. 
The equations of {\ECGR} are rewritten such that they formally resemble those of Einstein General Relativity ({\EGR}); this allows ideas from post-Newtonian theory to be imported without \wording{essential change}. 
The equations of motion are then written both at first post-Newtonian (1PN) order and at 1.5PN order. 
{\ECGR}'s 1PN equations of motion are found to be those of a micropolar/Cosserat elastic medium, along with a decoupled evolution equation for non-classical, spin-related fields. 
It seems that a necessary condition for these results to hold is that one chooses the non-classical fields to scale with the speed of light in a certain empirically reasonable way. 
Finally, the 1.5PN equations give greater insight into the coupling between energy-momentum and spin within slowly moving, weakly gravitating matter.
\textcolor{black}{
Specifically, the weakly relativistic modifications to Cosserat theory involve a gravitational torque and an augmentation of the gravitational force due to a `dynamic mass moment density' with an accompanying `dynamic mass moment density flux',
and new forms of linear momentum density captured by a `dynamic mass density flux' and a `dynamic momentum density'.
}
\end{abstract}

\maketitle


\section{Introduction} 
\label{section:Introduction}

The theory of general relativity (GR) with spin and torsion was initially proposed by~\citet{cartan1923,cartan1924,cartan1925}. 
This \textit{Einstein-Cartan (EC) theory} of GR (also known as $U^4$ theory and Einstein--Cartan--Sciama--Kibble relativity) was rediscovered in the 1960s by~\citet{Kibble1961} and~\citet{Sciama1962,Sciama1964}, summarized by~\citet{Hehl1976},
and remains a viable theory of relativity~\citep{Trautman2006,Hehl2007}.
There has been recent interest in {\ECGR} because it can avoid the Big Bang singularity~\citep[e.g.,][]{Poplawski2010,Poplawski2012}. As \cite{Hehl1976} discuss, {\ECGR} is a natural generalisation of general relativity to media where the quantum mechanical spin of matter is relevant.

In classical continuum mechanics,
a medium that supports spin is referred to as a `Cosserat micropolar medium',
in reference to the Cosserat brothers~\citep{cosserat1896,cosserat1907,cosserat1909} who inspired Cartan's work on torsion and spin.
The Cosserat equations are discussed by~\citet{truesdell1960},
and more modern descriptions of the theory are presented by~\citet{malvern1969} and~\citet{nowacki1986};
the latter also discusses the physical properties of micropolar media. \citet{maugin2010mechanics} provide a \wording{full} overview of both Cosserat theory and its generalisations, with \citet{maugin2010generalized} providing a useful conceptual introduction. 

Continuum mechanics is a coarse-grained theory of the atom-scale interactions occurring within matter. As discussed by \cite{maugin2010generalized}, if one assumes that the traction exerted on a facet cut in the solid depends only on the local unit normal, \edit{then} \textit{Cauchy's theorem} holds, and one may write the internal body forces as the divergence of a tensor: the \textit{Cauchy stress}. If, in addition, there are no applied couples in both volume and surface -- meaning essentially that one can assume that the particles of the medium interact by central forces -- then the Cauchy stress is symmetric. Such a medium is described by the well-known equations of motion
\begin{equation}
\partial_t \rho+\nabla_j (\rho\,v^j)=0
\quad,
\end{equation}
\begin{equation}
\label{eq:momConsInitialBasicGen}
\partial_t (\rho\,v_i)+\nabla_j (\rho\,v_i\,v^j-\sigma_i{}^j)
=  f_{i}
\quad,
\end{equation}
where \( \rho \) is the mass-density, \( v^{i} \) the material velocity, \( ?\sigma_{i}^{j}? \) the Cauchy stress, and \( f_{i} \) some external body force per unit mass. Given a constitutive theory relating the Cauchy stress to the deformation of the medium, these equations may be solved for the \wording{motion} of the medium.

But on a small enough scale the forces between particles are not central, with spin--spin interactions between atoms being perhaps the most obvious example. This motivates the Cosserat brothers' more general coarse-grained theory, where material points are considered to be defined not just by their mass-density, but also by a \textit{spin-density}. Within that theory the Cauchy stress is no longer required to be symmetric, and the equations above are augmented by the new equation
\begin{equation}
\label{eq:spinConsInitialBasicGen}
\partial_t \theta_{ij}+\nabla_k (\theta_{ij}\,v^k-\coupStress_{ij}{}^k)= \mbox{}-(\sigma_{ij}-\sigma_{ji})
+\Psi_{ij}
\quad,
\end{equation}
where the antisymmetric tensor \( \theta_{ij} \) gives the local spin density that is sourced by the \textit{material couple stress} \( ?\coupStress_{ij}^{k}? \) and whose evolution is forced both by an external torque \( \Psi_{ij} \) and by the asymmetry of the Cauchy stress. 
Again, after imposing suitable constitutive theory on \( ?\coupStress_{ij}^{k}? \) 
one could solve these equations for both the motion and the spin density.

The respective motivations for Cosserat continuum mechanics and {\ECGR} are in some sense the same: if one considers small enough scales, the spin of matter must impart some further structure into the coarse-grained theories describing bulk matter. Thus, classical continuum mechanics is `promoted' to Cosserat theory, and torsionless Einstein General Relativity ({\EGR}) to torsionful {\ECGR}. Moreover, just as there are myriad extensions of {\EGR}, continuum mechanics may also be augmented in several ways \citep[e.g.,][]{maugin2010mechanics}. 

Can the analogy be extended further? The equations of classical continuum mechanics are well-known to be derivable as the weak-field limit of {\EGR} \citep[e.g.,][]{poisson2014gravity}. Is the same true of Cosserat theory and {\ECGR}? This paper aims to show that, subject to certain assumptions about the spin and energy-momentum tensors, the equations of Cosserat theory may indeed be derived as the weak-field limit of Einstein--Cartan relativity. An additional aim is to study the coupling between energy-momentum and spin for slowly-moving, weakly self-gravitating matter; to that end, we also derive equations of motion in a limit where the fields are a little less weak.

We begin in Section~\ref{section:Einstein-Cartan general relativity} by reviewing {\ECGR}, discussing in particular how it may be rewritten without approximation to resemble {\EGR} rather closely. Then in Section~\ref{sec:anatomy} we discuss the structure of {\ECGR}'s energy-momentum and spin tensors and postulate a particular post-Newtonian scaling for some of their components. The weak-field equations of the theory are derived in Section~\ref{section:classical limit}; since we aim to derive the kinematic equations given just above, we impose no particular constitutive relation. In Section~\ref{section:onePointFive} we present the `slightly less weak field' equations, before discussing our results in Section~\ref{section:discussion}.

\section{Overview/review of Einstein--Cartan gravity}
\label{section:Einstein-Cartan general relativity}

{\ECGR} describes a spacetime \wording{containing} both energy-momentum -- represented by the asymmetric tensor \( T_{\mu\nu} \) -- and spin \( ?M_{\mu\nu}^{\sigma}? \). For that reason, {\ECGR} has a richer geometrical structure than \EGR.
\subsection{Geometry}

The spacetime of {\ECGR} is a four-dimensional Riemannian manifold endowed with a metric tensor \( g_{\mu\nu} \) and a metric-compatible affine connection \( \Gamma^{\sigma}_{\mu\nu} \). In contrast to {\EGR}, the connection need not be symmetric in its lower indices; its asymmetry is measured by the \textit{torsion tensor}, which we define as
\begin{equation}
\label{eq:torsiongr}
    ?S_{\mu\nu}^{\sigma}?
    =
    \Gamma^{\sigma}_{\mu\nu}-\Gamma^{\sigma}_{\nu\mu}
    =
    \mbox{}-?S_{\nu\mu}^{\sigma}? \quad.
\end{equation}
To make closer contact with {\EGR}, we may write the connection as 
\begin{align}
\label{eq:connectionGR}
  \Gamma^{\sigma}_{\mu\nu}
  =
  \hat{\Gamma}^{\sigma}_{\mu\nu}
  +
  ?K_{\mu\nu}^{\sigma}? \quad,
\end{align}
where 
\begin{align}
  \Gamhat^{\sigma}_{\mu\nu}
  =
  \tfrac{1}{2} 
  g^{\sigma\beta}
  \left(
     \partial_{\mu}g_{\beta \nu}
    +\partial_{\nu}g_{\mu \beta}
    -\partial_\beta g_{\mu\nu}
  \right)
  =
  \Gamhat^{\sigma}_{\nu\mu}
\end{align}
are the standard \textit{Christoffel symbols}, and
\begin{align}
\label{eq:contortion}
  K^{\mu\nu\sigma}
  =
  \tfrac{1}{2}
  (
     S^{\mu\nu\sigma}
    +S^{\sigma\mu\nu}
    +S^{\sigma\nu\mu}
  )
  =
  \mbox{}-K^{\mu\sigma\nu}
\end{align}
is the \textit{contortion tensor}. Note that \( K \)'s antisymmetry is on its second and third indices and that its first two indices need to have no particular symmetry. 
Expressions involving permutations of indices as in eq.~\eqref{eq:contortion} will arise several times, so we define the shorthand
\begin{align}
\label{eq:permutationShorthand}
  \widetilde{A}^{\mu\nu\sigma}
  =
  A^{\mu\nu\sigma}
  +
  A^{\sigma\mu\nu}
  +
  A^{\sigma\nu\mu} \quad,
\end{align}
for an arbitrary rank-3 tensor \( A^{\mu\nu\sigma} \).
Finally, for later reference, we define a third measure of torsion: the \textit{modified torsion tensor}~\citep[][eqn.~2.3; also referred to as the \textit{Palatini torsion tensor}]{Hehl1976}
\begin{equation}
\label{eq:Cartangr}
C{}_{\mu\nu}{}^\sigma= S{}_{\mu\nu}{}^\sigma+S{}_{\alpha \mu}{}^\alpha\delta_\nu{}^\sigma-S{}_{\alpha\nu}{}^\alpha\delta_\mu{}^\sigma
=\mbox{}-C{}_{\nu\mu}{}^\sigma
\quad .
\end{equation}

From the connection~\eqref{eq:connectionGR} we may define the \textit{covariant derivative} \( \nabla \), which acts on tensors according to
\begin{align}
  \nabla_{\sigma}?A_{\mu}^{\nu}?
  =
  \partial_{\sigma}?A_{\mu}^{\nu}?
  +
  \Gamma^{\nu}_{\sigma\alpha}?A_{\mu}^{\alpha}?
  -
  \Gamma^{\alpha}_{\sigma\mu}?A_{\alpha}^{\nu}? \quad.
\end{align}
Two other useful derivatives are the \textit{modified covariant derivative} \( \covst{} \),
\begin{align}
  \covst{\sigma}?A_{\mu}^{\nu}? 
  = 
  \nabla_{\sigma}?A_{\mu}^{\nu}? 
  + 
  ?S_{\sigma\alpha}^{\alpha}?
  ?A_{\mu}^{\nu}? \quad;
\end{align}
and the \textit{metric covariant derivative} \( \covmet{} \),
\begin{align}
  \covmet{\sigma}?A_{\mu}^{\nu}?
  =
  \partial_{\sigma}?A_{\mu}^{\nu}?
  +
    \hat{\Gamma}^{\nu}_{\sigma\alpha}?A_{\mu}^{\alpha}?
-
    \hat{\Gamma}^{\alpha}_{\sigma\mu}?A_{\alpha}^{\nu}? \quad.
\end{align}

The \textit{Riemann} or \textit{curvature tensor}\footnote{We use the definition of~\citet{Hehl1976} for the Riemann tensor,
which has the same placement of indices as~\citet{wald1984} but opposite sign.} is defined in terms of the connection coefficients~\eqref{eq:connectionGR} as
\begin{equation}
R{}_{\mu\nu\alpha}{}^\sigma = \partial_\mu\Gamma_{\nu\alpha}^\sigma-\partial_\nu\Gamma_{\mu\alpha}^\sigma
+\Gamma_{\mu\beta}^\sigma\,\Gamma_{\nu\alpha}^\beta
-\Gamma_{\nu\beta}^\sigma\,\Gamma_{\mu\alpha}^\beta
\quad .
\label{eq:Riemannngr}
\end{equation}
Unlike in {\EGR}, in {\ECGR} we do not have the symmetry~$R{}_{\mu\nu\alpha\sigma}=R{}_{\alpha\sigma\mu\nu}$\,. Nevertheless, the curvature tensor is still antisymmetric in both its first and its last two indices:
\begin{subequations}
\begin{align}
R{}_{\mu\nu\alpha\sigma} =-\mbox{}R{}_{\nu\mu\alpha\sigma} \quad,
\label{eq:Riemannngrasym1}
\\
R{}_{\mu\nu\alpha\sigma} =-\mbox{}R{}_{\mu\nu\sigma\alpha} \quad.
\label{eq:Riemannngrasym2}
\end{align}
\end{subequations}
As noted by \cite{Hehl1976}, even in the presence of torsion the \textit{Ricci tensor}~$R_{\mu\nu}$ therefore remains the only ``essential contraction" of the Riemann tensor:
\begin{equation}
\begin{split} 
R{}_{\mu\nu} \ = & \  R{}_{\sigma\mu\nu}{}^\sigma
\\ = & \ \partial_\sigma\Gamma_{\mu\nu}^\sigma-\partial_\mu\Gamma_{\sigma\nu}^\sigma
+\Gamma_{\sigma\beta}^\sigma\,\Gamma_{\mu\nu}^\beta
-\Gamma_{\mu\beta}^\sigma\,\Gamma_{\sigma\nu}^\beta
\quad .
\end{split}
\label{eq:riccigr}
\end{equation}
Defining the \textit{Ricci scalar} by
\begin{equation}
\label{eq:ricciscalargr}
    R=R_{\mu\nu}\,g^{\mu\nu}
    \quad ,
\end{equation}
we may then write the \textit{Einstein tensor} as
\begin{equation}
G_{\mu\nu}= R_{\mu\nu}-\tfrac{1}{2}R\,g_{\mu\nu} \quad.
\label{eq:Einsteingr}
\end{equation}
It is a distinguishing feature of {\ECGR} that the Ricci and Einstein tensors are no longer symmetric.

Just like the connection~\eqref{eq:connectionGR}, the Ricci and Einstein tensors may both be written as the sum of two terms: a `metric part' having no algebraic dependence on torsion, and a `non-metric part' that does have such a dependence and that vanishes in torsion's absence. 
Therefore, we express the Ricci tensor~\eqref{eq:riccigr} in the form
\begin{equation}
    R{}_{\mu\nu} =\hat{R}{}_{\mu\nu} +\mathring{R}{}_{\mu\nu} 
    \quad ,
\end{equation}
with the metric part given by
\begin{equation}
\hat{R}{}_{\mu\nu} =
\partial_\sigma\hat{\Gamma}_{\mu\nu}^\sigma
-\partial_\mu\hat{\Gamma}_{\sigma\nu}^\sigma
+\hat{\Gamma}_{\sigma\beta}^\sigma\, \hat{\Gamma}_{\mu\nu}^\beta
-\hat{\Gamma}_{\mu\beta}^\sigma\, \hat{\Gamma}_{\sigma\nu}^\beta
\quad ,
\end{equation}
and the part associated with torsion determined by
\begin{equation}
\begin{split} 
&
\mathring{R}{}_{\mu\nu} =
\nabla_\sigma K_{\mu\nu}{}^\sigma
-\nabla_\mu K_{\sigma\nu}{}^\sigma
-S_{\mu\sigma}{}^\alpha\, K_{\alpha\nu}{}^\sigma
\\ & \mbox{} \qquad\qquad
+K_{\mu\beta}{}^\sigma\, K_{\sigma\nu}{}^\beta
-K_{\sigma\beta}{}^\sigma\, K_{\mu\nu}{}^\beta
\quad .
\end{split}
\end{equation}
If we then define a symmetric second-rank tensor%
\footnote{For later convenience we take the argument of the LHS to be \( ?C_{\mu\nu}^{\sigma}? \) even while we write the RHS in terms of \( K^{\mu\sigma\nu} \).}
\begin{align}
  \SQuad_{\mu\nu}(C)
  &=
  ?K^{\alpha}_{\alpha\mu}?\,
  ?K^{\beta}_{\beta\nu}?
  -
  ?K_{\alpha\mu}^{\beta}? \,
  ?K_{\beta\nu}^{\alpha}?
  \nonumber\\
  &\quad
  -
  \tfrac{1}{2}\,g_{\mu\nu}
  (
    ?K^{\alpha}_{\alpha\sigma}?\,
    ?K^{\beta}_{\beta}^{\sigma}?
    +
    ?K_{\alpha\beta}^{\sigma}? \,
    ?K_{\sigma}^{\alpha\beta}? 
  )
\end{align}
that depends on \( C \) quadratically 
\citep[c.f.,][eq.~32]{battista2021},
and if we also recall the shorthand defined in eq.~\eqref{eq:permutationShorthand},
then it follows that the Einstein tensor~\eqref{eq:Einsteingr} may be expressed as
\begin{equation}
    G{}_{\mu\nu} =\hat{G}{}_{\mu\nu} +\mathring{G}{}_{\mu\nu} 
    \quad ,
\end{equation}
with the familiar metric part
\begin{equation}
\label{eq:einsteinMetric}
\hat{G}_{\mu\nu} = \hat{R}_{\mu\nu} - \tfrac{1}{2}\hat{R}\,g_{\mu\nu} \quad,
\end{equation}
and the torsionful part
\begin{align}
\label{eq:einsteinNonMetric}
  \Gcirc_{\mu\nu}
  =
  \tfrac{1}{2}\covst{\sigma}\Ct{}_{\mu\nu}{}^{\sigma}
  -
  \SQuad_{\mu\nu}(C) \quad.
\end{align} 


\subsection{Field equations}

The EC field equations may be obtained by considering the action~\citep[e.g.,][]{Hehl1973,Hehl1974,Hehl1976}
\begin{equation}
I=\int (\mathscr{L}_G+\mathscr{L}_M)\sqrt{-g}\,d^4 x \quad,
\label{eq:EHaction}
\end{equation}
where~$g$ denotes the determinant of the metric tensor~$g_{\mu\nu}$\,.
The Lagrangian density of the gravitational field is defined in terms of the Ricci scalar~\eqref{eq:ricciscalargr} by
\begin{equation}
    \mathscr{L}_G=\frac{1}{2\kappa}\,R
    \quad ,
\end{equation}
where the constant~$\kappa$ is determined in terms of Newton's gravitational constant, $G$\,, and the speed of light in vacuum, $c$\,, by~$\kappa=8\pi G/c^4$\,.
The Lagrangian density of matter is denoted by~$\mathscr{L}_M$\,.
Both Lagrangian densities are considered to be functions of the metric tensor~$g_{\mu\nu}$ and -- depending on algebraic convenience -- the torsion~$S_{\mu\nu}{}^{\sigma}$ or the contortion~$K_{\mu\nu}{}^{\sigma}$\,~\citep{Hehl1976}.

Variation of the matter Lagrangian density with respect to the metric yields the \textit{metric stress-energy tensor}
\begin{equation}
    \hat{T}^{\mu\nu}=\frac{\mbox{}-2}{\sqrt{-g}}\frac{\partial(\mathscr{L}_M\,\sqrt{-g})}{\partial g_{\mu\nu}}
    =\hat{T}^{\nu\mu}
    \quad ,
\end{equation}
while variation of the matter Lagrangian density with respect to contortion yields the \textit{spin tensor}
\begin{equation}
    M{}_\sigma{}^{\nu\mu}=\frac{\mbox{}-2}{\sqrt{-g}}\frac{\partial(\mathscr{L}_M\,\sqrt{-g})}{\partial K_{\mu\nu}{}^\sigma}
    =\mbox{}-M{}^{\nu}{}_\sigma{}^{\mu}
    \quad .
\end{equation}
These may be combined to give the \textit{canonical stress-energy tensor}~\citep[e.g.,][eqn.~3.8]{Hehl1976}
\begin{equation}
\label{eq:total stress}
    T^{\mu\nu}
    =
    \hat{T}^{\mu\nu}
    +
    \tfrac{1}{2}
    \covst{\sigma}
      \widetilde{M}^{\mu\nu\sigma}
\quad .
\end{equation}
Like the Ricci and Einstein tensors, \( T^{\mu\nu} \) is not symmetric in {\ECGR}.

Variation of the EC action~\eqref{eq:EHaction} now yields the field equations~\citep{Hehl1976}
\begin{subequations}
\label{eq:EC}
\begin{align}
  G_{\mu\nu}
  &=
  \kappa\,T_{\mu\nu} \quad,
  \label{eq:EC1}
  \\  
  C_{\mu\nu}{}^\sigma
  &=
  \kappa\,M_{\mu\nu}{}^\sigma \quad.
  \label{eq:EC2}
\end{align}
\end{subequations}
It is noteworthy that, while the Einstein Field Equation~\eqref{eq:EC1} (EFE) is a partial differential equation, the spin field eqn.~\eqref{eq:EC2} is an algebraic identity. This means that spin vanishes if and only if torsion does too.

\subsection{Dynamic equations}
\label{section:dynamic_equations}

The conservation laws for energy-momentum and spin are found by Noether's theorem \citep[e.g.,][]{Hehl1976}) to be
\begin{subequations}
\label{eq:ECFE}
\begin{align}
  \covst{\nu}T_\mu{}^\nu
  &=
  \tfrac{1}{2}M{}_{\sigma\alpha}{}^{\nu}\,R{}_{\mu\nu}{}^{\sigma\alpha}
  +S{}_{\mu\alpha}{}^\nu \,T_\nu{}^\alpha 
  \quad ,
  \label{eq:ECFE1p}
  \\
  \covst{\sigma} M_{\mu\nu}{}^\sigma
  &=
  T_{\mu\nu}-T_{\nu\mu}
  \quad.
  \label{eq:ECFE2p}  
\end{align}
\end{subequations}
Note that it is the canonical stress-energy \( T_{\mu\nu} \) that appears here, not its metric cousin \( \hat{T}_{\mu\nu} \). 
These identities are the `dynamic equations' of {\ECGR}~\citep[e.g.,][]{Hehl1986,lovelock1989}. As discussed by \cite{Hehl1986}, they are automatically compatible with the contracted Bianchi identities 
as long as the field equations~\eqref{eq:EC} are fulfilled. Thus, similarly to {\EGR}, {\ECGR} possesses ``automatic conservation of the source'' \citep[see, e.g.,][]{misner1973}.

\subsection{`Effective' equations of {\ECGR}}

In the absence of spin,
as in {\EGR}~\citep{einstein1918},
the spin field equation~\eqref{eq:EC2} forces torsion to vanish and the canonical- and metric stress-energy tensors to coincide. The field- and dynamic equations thus reduce to
\begin{equation}
\Ghat_{\mu\nu} = \kappa T_{\mu\nu}
\quad,
\label{eq:ECFE0NT}
\end{equation}
and
\begin{equation}
\covmet{\nu}?T_{\mu}^{\nu}?=0
\quad,
\label{eq:ECFE1NT}
\end{equation}
with
\begin{equation}
T_{\mu\nu}=T_{\nu\mu}
\quad.
\label{eq:ECFE2NT}
\end{equation}
In the absence of torsion, the stress-energy tensor is symmetric.

But even in the presence of spin, the equations of {\ECGR} may be rewritten without approximation so as to resemble those of {\EGR} rather closely \citep[see][]{Hehl1976}. To see this we may use the decomposition of the Einstein tensor into metric- and non-metric parts (eqs.~\ref{eq:einsteinMetric}, \ref{eq:einsteinNonMetric}) and the trivial identities
\begin{align}
  \Ct{}_{(\mu\nu)}{}^{\sigma}
  &=
  C{}^{\sigma}{}_{(\mu\nu)}\quad,
  \\
  \Ct{}_{[\mu\nu]}{}^{\sigma}
  &=
  C{}_{\mu\nu}{}^{\sigma} \quad.
\end{align}
If we then split the EFE~\eqref{eq:EC1} into symmetric and antisymmetric parts, we find that
\begin{subequations}
\begin{align}
  &
  \Ghat_{\mu\nu}
  +
  \tfrac{1}{2}\covst{\sigma}C{}^{\sigma}{}_{(\mu\nu)}
  -
  \SQuad_{\mu\nu}(C)
  =
  \kappa\, T_{(\mu\nu)} \quad,
  \\
  &
  \tfrac{1}{2}\covst{\sigma}C{}_{\mu\nu}{}^{\sigma}
  =
  \kappa\, T_{[\mu\nu]} \quad.
\end{align}
\end{subequations}
Now we use the spin-field equation~\eqref{eq:EC2} to substitute spin for torsion everywhere (including within~\( \covst{\sigma} \)):
\begin{subequations}
\begin{align}
  & \!\!\!\!\!
  \Ghat_{\mu\nu}
  =
  \kappa\, T_{(\mu\nu)}
  -
  \tfrac{1}{2}\kappa\covst{\sigma}M{}^{\sigma}{}_{(\mu\nu)}
  +
  \kappa^{2} \SQuad_{\mu\nu}(M) \quad,
  \label{eq:effectiveEFEInterm}
  \\
  &
  \covst{\sigma}M{}_{\mu\nu}{}^{\sigma}
  =
  2\, T_{[\mu\nu]} \quad.
\end{align}
\end{subequations}
The second of these equations is a redundant repetition of the conservation law for spin~\eqref{eq:ECFE2p}. 

Finally, define the symmetric \textit{effective energy--momentum tensor}
\begin{align}
\label{eq:Theta_defn}
  \Theta_{\mu\nu}
  =
  T_{(\mu\nu)}
  -
  \tfrac{1}{2}\covst{\sigma}M{}^{\sigma}{}_{(\mu\nu)}
  +
  \kappa\, \SQuad_{\mu\nu}(M) \quad,
\end{align}
which may alternatively be written as
\begin{align}
  \Theta_{\mu\nu}
  =
  \hat{T}_{\mu\nu}
  +
  \kappa\, \SQuad_{\mu\nu}(M) \quad,
\end{align}
using eqs.~(\ref{eq:total stress}, \ref{eq:EC2}).
We will generally use the first alternative because this paper's focus is on the (weak-field) kinematics of the physically meaningful matter fields of the canonical stress-energy tensor. In any case, the symmetrized effective EFE~\eqref{eq:effectiveEFEInterm} thus takes the familiar form
\begin{align}
  \Ghat_{\mu\nu}
  =
  \kappa\, \Theta_{\mu\nu} \quad.
\end{align}
The metric Einstein tensor is conserved by the metric covariant derivative, and therefore so is \( \Theta_{\mu\nu} \):
\begin{align}
  \covmet{\nu}\Theta^{\mu\nu} = 0 \quad.
\end{align}
(Presumably, this conservation law follows from algebraic manipulation of the field equations~\eqref{eq:EC} and the dynamic equations~\eqref{eq:ECFE}, but it seems more straightforward just to appeal to \( \Ghat_{\mu\nu} \)'s properties.) 

In summary, the effective equations of {\ECGR} are
\begin{align}
  &
  \Ghat_{\mu\nu} = \kappa\,\Theta_{\mu\nu} \quad,
  \label{eq:effectiveEFE}\\
  &
  \covmet{\nu}\Theta^{\mu\nu} = 0 \quad ,
  \label{eq:effectiveConservation}\\
  &
  \covst{\sigma}M{}_{\mu\nu}{}^{\sigma}
  =
  2 \,T_{[\mu\nu]} \quad,
  \label{eq:spinConservation}
\end{align}
with all instances of torsion within \( \covst{\sigma} \) understood to have been replaced by spin using eq.~\eqref{eq:EC2}. The first two of these equations are algebraically identical to standard {\EGR}, albeit with a differently defined stress-energy tensor. 
\wording{However, they do not obviously decouple from the spin conservation equation, so the similarity is just formal.}

\section{Anatomy of the stress-energy and spin tensors}
\label{sec:anatomy}

Before deriving the equations of weak-field {\ECGR}, we should study the structure of the stress-energy and spin tensors more closely. 
We use an index~$0$ to denote the time coordinate, $x^0=c\,t$\,, and lowercase Roman
indices to denote three spatial coordinates, $\{x^i\}$\,.
The speed of light, $c$\,, is introduced as a scaling parameter
that enables us to keep track of and order various terms in powers of~$c$\,.
The structure of the scale is set by the rest energy, $T_{00}=\rho\,c^2$.

We assume that in an \textit{instantaneously co-moving Lorentz frame} (ICLF) the energy-momentum tensor takes the form
\begin{equation}
\begin{split}
    (\,T^{\mu\nu}\,) \ = & \ 
    \left(\begin{array}{cc}
    T^{00} & T^{0j}  \\
    T^{i0} & T^{ij} 
    \end{array}\right) 
    \\ \equiv & \ 
    \left(\begin{array}{cc}
    c^2\,\rho & q^j  \\
    p^i & \mbox{}-\sigma^{ij} 
    \end{array}\right) 
    \quad .
    \end{split}
\label{eq:Tans2}
\end{equation}
We have included a nonzero ICLF momentum-density \( p^{i} \) and mass-flux \( q^{i} \), as well as dropping the {\EGR} requirement that \( \sigma^{ij} \) be symmetric. We may derive the corresponding expression for \wording{slowly-moving matter} by Lorentz-boosting with a small velocity \( v^{i} \), where \( \|v\| \ll c \). Defining
\begin{align}
  \Sigma^{ij} 
  &=
  \sigma^{ij} - \rho\,v^{i} v^{j} \quad,
  \label{eq:Tdefinition}
\end{align}
the stress-energy tensor takes the generic form
\begin{equation}
    (\,T{}^\mu{}^\nu\,)=\left(\begin{array}{cc}
    c^2\,\rho & c\,\rho\,v^j+q^j  \\
    c\,\rho\,v^i+p^i & \mbox{}-\Sigma^{ij} + (p^{i}v^{j}+v^{i}q^{j}) /c
    \end{array}\right) 
    \quad ,
    \label{eq:Tmunu2}
\end{equation}
where we have included the leading-order term in each entry, as well as the term one power of \( c \) lower. 

We have chosen \( p^{i} \) and \( q^{i} \) to carry one less factor of \( c \) than \( \rho\, v^{i} \). This scaling seems empirically reasonable. At low enough velocities, an object's linear momentum is derived solely from its rest mass and velocity. But \( p^{i} \) and \( q^{i} \) represent contributions from spacetime itself, so they must enter at sub-leading order as in eq.~\eqref{eq:Tmunu2}.

As for the spin tensor, we take its ICLF components to be
\begin{equation}
\begin{split} 
\left(\,M^{\mu\nu}{}^{0}\,\right)
\ = & \ 
\left(\begin{array}{cc}
0 & M^{0j}{}^{0} \\
M^{i0}{}^{0} & M^{ij}{}^{0}
\end{array}\right)
    \\ \equiv & \ 
\left(\begin{array}{cc}
0 & \mbox{}-c\,\mu^j \\
c\,\mu^i & c\,\theta^{ij}
\end{array}\right)
\quad ,
\end{split} 
\label{eq:Man2}
\end{equation}
\begin{equation}
\begin{split} 
\left(\,M^{\mu\nu}{}^{k}\,\right)
\ = & \ 
\left(\begin{array}{cc}
0 & M^{0j}{}^{k} \\
M^{i0}{}^{k} & M^{ij}{}^{k}
\end{array}\right)
    \\ \equiv & \ 
\left(\begin{array}{cc}
0 & \massMom^j{}^k \\
\mbox{}-\massMom^i{}^k & \mbox{}-\coupStress^{ij}{}^{k}
\end{array}\right)
\quad .
\end{split} 
\label{eq:Mans2}
\end{equation}
Again, we have scaled the non-classical fields \( \mu^{i} \) and \( \massMom^{ik} \) to ensure that non-classical effects cannot enter at leading order. 
Lorentz-boosting once again with the small velocity \( v^{i} \), and defining (\cf eq.~\ref{eq:Tdefinition})
\begin{align}
  \MassMom^{ij}
  &=
  \massMom^{ij} - \mu^{i} v^{j} \quad ,
  \label{eq:definitionsPi}\\
  \CoupStress^{ijk}
  &=
  \coupStress^{ijk} - \theta^{ij}v^{k} \quad,
  \label{eq:definitionsLam}
\end{align}
the spin tensor takes the generic leading-order form
\begin{equation}
\left(\,M^{\mu\nu}{}^{0}\,\right)
 =
\left(\begin{array}{cc}
0 & \mbox{}-c\,\mu^j \\
c\,\mu^i & c\,\theta^{ij} 
\end{array}\right)
\quad ,
\label{eq:Mmunu02}
\end{equation}
and
\begin{equation}
\left(\,M^{\mu\nu}{}^{k}\,\right)=
\left(\begin{array}{cc}
0 & \mbox{}\MassMom^{jk} \\
\mbox{}-\MassMom^{ik} & \mbox{}-\CoupStress^{ijk} 
\end{array}\right)
\quad .
\label{eq:Mmunuk2}
\end{equation}
One may show that, as in eq.~\eqref{eq:Tmunu2}, non-classical effects correct \( \theta^{ij} \) and \( \CoupStress^{ijk} \) only at sub-leading order in \( c \).

Thus, the physical parameters of {\ECGR} are seen to include:
the mass density~$\rho$,
the dynamic mass density flux~$c\,q^i$,
the dynamic momentum density~$c\,p^i$,
the material stress tensor~$\sigma{}^{ij}$,
the dynamic mass moment density~$c\,\mu^i$,
the particle spin tensor~$\theta^{ij}$,
the dynamic mass moment density flux~$c\,\massMom^{ij}$,
and the material couple stress tensor~$\coupStress^{ij}{}^k$\,.
\wording{The justification for this nomenclature will be motivated later in Section~\ref{section:eqmon}.}
From a classical continuum mechanics perspective,
new terms in the stress-energy tensor are the dynamic mass density flux~$c\,q^i$
and the dynamic momentum density~$c\,p^i$\,, 
while the new spin terms involve the dynamic mass moment density~$c\,\mu^i$ and the dynamic mass moment density flux~$c\,\massMom^{ij}$\,.
The theory's physical parameters are summarized in Table~\ref{table:new quantities2}.

\begin{table}[t!]
    \centering
    \begin{tabular}{c|c}
      $\rho$   &  mass density  \\
      $v^i$   &  particle velocity  \\
      $\sigma^{ij}$   &  material stress tensor  \\
      $\theta^{ij}$   &  particle spin tensor  \\
      $\coupStress^{ijk}$   &  material couple stress tensor  \\ \\
      \hline \\
      $c\,q^i$   &  dynamic mass density flux \\
      $c\,p^i$   &  dynamic momentum density \\
      $c\,\mu^i$   &  dynamic mass moment density \\
      $c\,\massMom^{ij}$   &  dynamic mass moment density flux
    \end{tabular}
    \caption{Summary of the physical parameters encountered in the weak-field limit of {\ECGR}.
    The top set of parameters is familiar from classical continuum mechanics with spin,
    whereas the bottom set of parameters represents \wording{relativistic} phenomena.}
    \label{table:new quantities2}
\end{table}

\section{Weak-field equations}
\label{section:classical limit}

\subsection{Preliminaries}

We now seek leading-order weak-field approximations to equations~(\ref{eq:effectiveEFE},~\ref{eq:effectiveConservation},~\ref{eq:spinConservation}), that is, to write down the equations of motion at \textit{first post-Newtonian order}. A post-Newtonian (PN) expansion combines two formal perturbation expansions: one in powers of \( c \)
and another in powers of \( G \) \citep[e.g.,][]{poisson2014gravity}. PN theory may systematically be pushed to high order, but since here we only seek equations accurate to first (`1-PN') order, it is not necessary to \wording{go so systematically}. Rather, it will be sufficient just to carry out traditional, linearised gravity theory \citep[e.g.,][\S 5.5]{poisson2014gravity}, dropping terms higher than \( \order(G) \) and solving the effective EFE~\eqref{eq:effectiveEFE} to leading order in \( c \) in each of \( \Theta_{00} \), \( \Theta_{0j} \) and \( \Theta_{ij} \). This approach's consistency with formal PN theory follows from comments on p.~307 of \cite{poisson2014gravity}.

For weak gravitational fields~\citep[e.g.,][]{misner1973,wald1984,Carroll,schutz2009},
we express the metric tensor as
\begin{equation}
    g_{\mu\nu}=\eta_{\mu\nu}+h_{\mu\nu}
    \quad ,
\end{equation}
where \( \eta_{\mu\nu} \) is the standard Minkowski metric with signature \( (-+++) \), and~$h_{\mu\nu}=h_{\nu\mu}$ denote small perturbations in the metric.
The linearised metric Einstein tensor is \citep[e.g.,][eq.~7.7]{Carroll}
\begin{align}
  \Ghat_{\mu\nu}
  &=
  \tfrac{1}{2}
  \brak{
   \partial_{\sigma}\partial_{\nu}?h^{\sigma}_{\mu}?  
  +\partial_{\sigma}\partial_{\mu}?h^{\sigma}_{\nu}?
  -\partial_{\mu}\partial_{\nu}h
  \mySplitLongBrackets{\qquad}
  \mbox{}-\square h_{\mu\nu}
  -\eta_{\mu\nu}\partial_{\lambda}\partial_{\rho}h^{\lambda\sigma}
  +\eta_{\mu\nu}\square h
  } \quad ,
\end{align}
where~$\square$ denotes the d'Alembertian~$h^{\mu\nu}\partial_{\mu}\partial_{\nu}$\,,
and the linearised Christoffel symbols are
\begin{align}
  \Gamhat^{\sigma}_{\mu\nu}
  =
  \tfrac{1}{2} 
  \eta^{\sigma\beta}
  \left(
     \partial_{\mu}h_{\beta \nu}
    +\partial_{\nu}h_{\mu \beta}
    -\partial_{\beta}h_{\mu\nu}
  \right) \quad.
\end{align}
As discussed in Appendix~\ref{sec:weakSolution}, \( h_{\mu\nu} \) scales like \( c^{-2} \) at most. 
The Christoffel symbols inherit this scaling, and since \( M{}^{\mu\nu\sigma} \) scales like \( c \) at most (eqs.~\ref{eq:Mmunu02},~\ref{eq:Mmunuk2}) it follows that any product of Christoffel symbols and spin is of order \( c^{-1} \). This will simplify the upcoming algebra.

\subsection{Reduction to linearised {\EGR}}

To leading order in \( G \), one may write the effective energy--momentum tensor~\eqref{eq:Theta_defn} as
\begin{align}
\label{eq:ThetaDefn}
  \Theta_{\mu\nu}
  =
  T_{(\mu\nu)}
  -
  \tfrac{1}{2}\covmet{\sigma}M{}^{\sigma}{}_{(\mu\nu)} \quad,
\end{align}
dropping \( \covst{} \)'s torsionful parts and \( \SQuad(M) \).
Additionally, the \wording{lowest leading order} in \( T_{(\mu\nu)} \) is \( \Sigma_{(ij)} \sim \order(c^{0}) \). Therefore, we may disregard the \( \order(c^{-1}) \) spin--Christoffel cross terms to leave
\begin{align}
\label{eq:Theta_PN}
  \Theta_{\mu\nu}
  =
  T_{(\mu\nu)}
  -
  \tfrac{1}{2}\partial_{\sigma}M{}^{\sigma}{}_{(\mu\nu)} \quad.
\end{align}
Applying analogous reasoning to the spin conservation equation~\eqref{eq:spinConservation} reduces it to
\begin{align}
\label{eq:spinConservationPN}
  \partial_{\sigma}M{}_{\mu\nu}{}^{\sigma}
  =
  2 \,T_{[\mu\nu]} \quad,
\end{align}
having noted in particular that \( T_{[\mu\nu]} \sim \order(c^{0}) \).

Thus our weak-field approximation to equations~(\ref{eq:effectiveEFE},~\ref{eq:effectiveConservation},~\ref{eq:spinConservation}) reduces to two decoupled problems. On the one hand we have the spin conservation equation~\eqref{eq:spinConservationPN} which has no algebraic dependence on the metric. On the other we have a standard problem of linearised {\EGR},
\begin{align}
  &
  \Ghat_{\mu\nu} = \kappa\,\Theta_{\mu\nu} \quad,
  \label{eq:effectiveEFE_weak}\\
  &
  \covmet{\nu}\Theta^{\mu\nu} = 0
  \label{eq:effectiveConservation_weak} \quad,
\end{align}
with effective stress~\eqref{eq:Theta_PN} given explicitly by
\begin{equation}
    (\,\Theta^{\mu\nu}\,)
    =
    \left(\begin{array}{cc}
    c^2\,\rho & 
    c\,(\rho\,v^{j}-\tfrac{1}{2}\,\partial_{k}\theta^{kj})  \\
    c\,(\rho\,v^{i}-\tfrac{1}{2}\,\partial_{k}\theta^{ki})  & 
    \partial_{k}\CoupStress^{k(ij)}-\Sigma^{(ij)}
    \end{array}\right) 
    \label{eq:Thmunu2}
\end{equation}
to leading order.

\wording{This decoupling} simplifies the process of gauge-fixing. Linearised {\EGR}'s well-known 
invariance under the transformations
\begin{align}
\label{eq:gaugeLinearisedNormal}
  h_{\mu\nu}
  \rightarrow
  h_{\mu\nu}
  +
  2\,\partial_{(\mu}\xi_{\nu)}
\end{align}
for arbitrary \( \xi_{\mu}(x) \) is not generally preserved within linearised torsionful theories. Analogous transformations do exist within such theories, but they tend to take a much more complicated form \citep[e.g.,][]{Barrientos2017,Barrientos2019}.
In this connection, we remark that \citet{battista2021} chose a specific constitutive relation for the spin (that of the \textit{Weyssenhof fluid}) to be able to treat the modified EFEs using the `standard' gauge-freedom of eq.~\eqref{eq:gaugeLinearisedNormal} in their study of gravitational waves with \wording{arbitrarily large torsion}. Within this paper we may make full use of the freedom represented by eq.~\eqref{eq:gaugeLinearisedNormal} because the weak-field approximation means that \wording{the spin does not see the metric}.

In fact, given the existence of extensive literature on linearised {\EGR}, at this point the problem is essentially solved. In eq.~\eqref{eq:Thmunu2} we have \( \Theta^{00} \) to \( \order(c^{2}) \), \( \Theta^{i0} \) to \( \order(c) \) and \( \Theta^{ij} \) to \( \order(1) \). As shown in \cite[][\S 5.5.7,~\S 7.3.2]{poisson2014gravity}, this is sufficient to write down the 1PN equations of energy-momentum conservation. All that is different from {\EGR} is the \textit{form} of the energy-momentum tensor, with the 3-momentum and 3-stress each \wording{possessing} spin-related \wording{corrections}. For those interested in our choice of gauge, we have included a full solution of eqs.~(\ref{eq:effectiveEFE_weak},~\ref{eq:effectiveConservation_weak}) in Appendix~\ref{sec:weakSolution}, but one can equally proceed straight to Section~\ref{section:eqmon}.

\subsection{Weak-field limit of the dynamic equations}
\label{section:eqmon}

From \cite[][eqs.~7.55-7.58]{poisson2014gravity} we find that 1PN conservation of \( \Theta_{\mu\nu} \) gives both standard conservation of mass,
\begin{align}
  \partial_{t}\rho
  +
  \partial_{k}(\rho\,v^{k})
  = 0 \quad,
\end{align} 
and a form of 3-momentum conservation,
\begin{equation}
\label{eq:threeMomentumConsWeak}
  \partial_{t}(
    \rho\,v^{i} 
    - 
    \tfrac{1}{2}\,\partial_{k}\theta^{ki}
  )
  +
  \partial_{j}(
    \partial_{k}\CoupStress^{k(ij)}
    -
    \Sigma^{(ij)}
  )
  =
  \mbox{}-\rho\,\partial_{i}V \quad.
\end{equation}
The field \( V \) is the standard Newtonian potential satisfying
\begin{align}
  \nabla^{2}V = 4 \pi G \rho \quad.
\end{align}

We find the equations of the spin degrees of freedom by taking the \( (i,0) \) and \( (i,j) \) parts of eq.~\eqref{eq:spinConservationPN}. This gives
\begin{align}
  \partial_{t}\mu^{i}
  -
  \partial_{k}\MassMom^{ik}
  =
  p^{i} - q^{i}
\end{align}
and
\begin{align}
\label{eq:threeSpinConsWeak}
  \partial_{t}\theta^{ij}
  -
  \partial_{k}\CoupStress^{ijk}
  =
  \mbox{}-2\,\Sigma^{[ij]} \quad.
\end{align}
The leading balance is at \( \order(c^{0}) \) in both cases, so the connection does indeed not contribute.

The last step is to use the equation for \( \theta^{ij} \) to bring the equation for \( \rho\, v^{i} \) into a more familiar form. We rewrite eq.~\eqref{eq:threeMomentumConsWeak} as
\begin{align}
\label{eq:threeMomentumConsWeak_interm}
  &
  \partial_{t}(
    \rho\,v^{i} 
  )
  -
  \partial_{j}
    \Sigma^{(ij)}
  =
  -\rho\,\partial^{i}V
  \nonumber\\ &
  \quad \mbox{} 
  +\partial_{k}(
    \tfrac{1}{2}\,\partial_{t}\theta^{ki}
    -
    \partial_{j}\CoupStress^{k(ij)}
  ) \quad,
\end{align}
then substitute eq.~\eqref{eq:threeSpinConsWeak} into the RHS of eq.~\eqref{eq:threeMomentumConsWeak_interm}. Given that \( \partial_{j}\partial_{k}\CoupStress^{kji} \) vanishes identically, it follows that
\begin{align}
  \partial_{t}(
    \rho\,v_{i} 
  )
  -
  \partial_{j}
    ?\Sigma_{i}^{j}?
  =
  \mbox{}-\rho\,\partial_{i}V \quad.
\end{align}

In summary,
using definitions~\eqref{eq:Tdefinition}, \eqref{eq:definitionsPi} and~\eqref{eq:definitionsLam},
the 1PN equations of Einstein--Cartan relativity are:
\begin{subequations}
\label{eq:dynamicWeakFinal}
\begin{align}
  &
  \partial_{t}\rho
  +
  \partial_{k}(\rho\,v^{k})
  = 0 \quad,
  \\
  &
  \partial_{t}(
    \rho\,v_{i} 
  )
  -
  \partial_{j}(?\sigma_{i}^{j}? - \rho\,v_{i} v^{j})
  =
  \mbox{}-\rho\,\partial_{i}V \quad,
  \label{eq:momDynamicWeakFinal}\\
  &
  \nabla^{2}V = 4\pi G \rho \quad,
  \\
   &
  \partial_{t}\theta_{ij}
  -
  \partial_{k}(?\coupStress_{ij}^{k}? - \theta_{ij}v^{k})
  =
  \mbox{}-(\sigma_{ij}-\sigma_{ji}) \quad,
  \label{eq:spinDynamicWeakFinal}
  \\
  &
  \partial_{t}\mu_{i}
  -
  \partial_{j}(?\massMom_{i}^{j}? - \mu_{i} v^{j})
  =
  p_{i} - q_{i}  \quad .
  \label{eq:spinWeirdDynamicWeakFinal}
\end{align}
\end{subequations}
\textcolor{black}{
Note that the only Christoffel symbols required for this result were
\begin{align}
  \hat{\Gamma}^{i}_{00} = c^{-2}\partial^{i}V + \order\brak{c^{-3}} ,
\end{align}
while we only needed to calculate the \wording{Newtonian part} of the metric:
\begin{align}
  \mathrm{d}s^{2}
  = &
  \brak{-1 + 2c^{-2}V}c^{2}\mathrm{d}t^{2}
  \nonumber
  \\
  &
  +\brak{1 - 2c^{-2}V}\,\delta_{ij}\,\mathrm{d}x^{i}\mathrm{d}x^{j}
  \quad .
\end{align}
}

In the first three of eqs.~\eqref{eq:dynamicWeakFinal} we see `classical' self-gravitating continuum mechanics, albeit with a stress tensor that need not be symmetric. Moreover, our solution of the EFEs has produced in eq.~\eqref{eq:momDynamicWeakFinal} a body force that is specifically gravitational (\cf eq.~\ref{eq:momConsInitialBasicGen}). The fourth equation adds Cosserat-style spin degrees of freedom that couple algebraically to the classical degrees of freedom through \( v^{i} \). Note that this equation is driven by the asymmetry of the stress tensor alone, with the external torque \( \Psi_{ij} \) vanishing at this order (\cf eq.~\ref{eq:spinConsInitialBasicGen}). Finally, we find an evolution equation involving the \wording{wholly non-classical} fields \( \mu^{i} \), \( \massMom^{ij} \), \( p^{i} \) and \( q^{i} \). It is this equation that motivates our earlier nomenclature: \( c\,\mu^{i} \) a `mass moment density' and \( c\,\massMom^{ij} \) the corresponding flux. Eq.~\eqref{eq:spinWeirdDynamicWeakFinal} also couples algebraically to the classical degrees of freedom through \( v^{i} \), although if~$\mu^i$\,, $\massMom^{ij}$\,, $p^i$\,, and~$q^i$ are zero initially then they remain so and~\eqref{eq:spinWeirdDynamicWeakFinal} is irrelevant.

\section{Slightly stronger fields}
\label{section:onePointFive}

We gain further insight into the interplay of spin and energy-momentum by writing
eqs.~(%
  \ref{eq:effectiveEFE}, 
  \ref{eq:effectiveConservation}, 
  \ref{eq:spinConservation}%
)
accurate to 1.5PN order.
It is easier to derive the 1.5PN equations than one might imagine, because \wording{no extra work is required regarding the EFEs:} as shown in Appendix~\ref{sec:OnePointFiveAlgebra}, we still only need the Newtonian potential \( V \). Moreover, we can continue to disregard the parts of \( \Theta^{\mu\nu} \) that explicitly feature torsion because they come with four extra factors of \( c^{-1} \) and are therefore suppressed at this PN order; eq.~\eqref{eq:ThetaDefn} is thus still a valid expression for the effective stress-energy tensor. All of this means that nothing from Appendix~\ref{sec:weakSolution} needs to be changed here. To obtain the 1.5PN equations, we just need to expand the dynamic equations 
~(%
  \ref{eq:effectiveConservation}, 
  \ref{eq:spinConservation}%
)
one order higher in \( c^{-1} \). This introduces no new concepts with respect to Section~\ref{section:classical limit}, just more algebra; therefore we consign most of the work to Appendix~\ref{sec:OnePointFiveAlgebra} while here we \wording{essentially just quote and describe the resulting equations}.

To 1.5PN order the spin tensor has components
\begin{subequations}
\begin{align}
  ?M^{0i0}?
  &=
  \mbox{}-c\,\mu^{i} - \theta^{ij}v^{j} + \order(c^{-1}) \quad,
  \\
  ?M^{ij0}?
  &=
  c\,\theta^{ij} + 2\,\mu^{[i}v^{j]} + \order(c^{-1})  \quad,
  \\
  ?M^{0ij}?
  &=
  \MassMom^{ij} - ?\CoupStress_{k}^{ij}?v^{k}/c + \order(c^{-2})\quad,
  \\
  ?M^{ijk}?
  &=
  \mbox{}-\CoupStress^{ijk} + 2\,v^{[i}\MassMom^{j]k}/c + \order(c^{-2}) \quad,
\end{align}
\end{subequations}
while the (symmetrised) canonical energy-momentum tensor is given by
\begin{subequations}
\begin{align}
  T^{00}
  &=
  c^{2}\rho 
  + 
  \order(c^{0}) \quad,
  \\
  T^{(0j)}
  &=
  c\,\rho\, v^{j}
  +
  \tfrac{1}{2}(p^{j}+q^{j})
  + 
  \order(c^{-1}) \quad,
  \\
  T^{(ij)}
  &=
  \mbox{}-\Sigma^{(ij)}
  +
  (p+q)^{(i}v^{j)}/c
  + 
  \order(c^{-2}) \quad.
\end{align}
\end{subequations}
Appendix~\ref{sec:OnePointFiveAlgebra} then shows that the 1.5PN equations of motion are 
(eqs.~%
  \ref{eq:muOnePointFive},
  \ref{eq:thetaOnePointFive},
  \ref{eq:rhoOnePointFive},
  \ref{eq:momOnePointFive}%
)
\begin{widetext}
\begin{subequations}
\label{eq:OnePointFiveFinal}
\begin{align}
  &
  \partial_t{\rho}
  +
  \partial_{j}(
    \rho\,v^{j} + q^{j}/c
  )
  =
  0
\quad,
  \label{eq:OnePointFiveFinal__Mass}\\
  &
  \partial_{t}(\rho\,v_{i} + p_{i}/c)
  -
  \partial_{j}[
    ?\sigma_{i}^{j}?
    -
    \rho\, v_{i}v^{j}
    -
    (
      p_{i}v^{j}
      + 
      v_{i}q^{j}
    )/c
  ]
  =
  -\rho\,\partial_{i}V
  -
  \mu^{j}\,\partial_{j}\partial_{i}V/c
\quad ,
  \label{eq:OnePointFiveFinal__Mom}\\
  &
  \nabla^{2}V = 4\pi G \rho \quad,
  \\
  &
  \partial_{t}[\theta_{ij} + (\mu_{i}v_{j}-\mu_{j}v_{i})/c)]
  -\partial_{k}\{?\coupStress_{ij}^{k}? - [\theta_{ij} + (\mu_{i}v_{j}-\mu_{j}v_{i})/c)]\,v^{k}
  -(v_{i}\,?\massMom_{j}^{k}?
  -v_{j}\,?\massMom_{i}^{k}?) 
  /c\}
  =
    \nonumber\\ & \qquad\qquad\qquad
    \mbox{}-(\sigma_{ij}-\sigma_{ji})
    +
    [(p_i-q_i)v_{j}-(p_j-q_j)v_{i}]/c
  -
  (\mu_{i\,}\partial_{j}V-\mu_{j\,}\partial_{i}V)/c
\quad ,
  \label{eq:OnePointFiveFinal__theta}\\  
  &
   \partial_{t}(\mu_{i}+\theta_{ij}\,v^{j}/c)
  -\partial_{j}[?\massMom_{i}^{j}? - \mu_{i} v^{j}
    -(?\coupStress_{ki}^{j}? - \theta_{ki}v^{j})v^{k}/c
   ]
  =
    p_{i}-q_{i}
    -
    (\sigma_{ij}-\sigma_{ji})v^{j}/c
  -
  ?\theta_{i}^{j}?\,\partial_{j}V/c
\quad.
  \label{eq:OnePointFiveFinal__mu}
\end{align}
\end{subequations}
\end{widetext}
\textcolor{black}{
We reiterate that the Newtonian potential \( V \) is the only \wording{extra} part of the metric required to derive these equations, just as in Section~\ref{section:eqmon}. Even upon the inclusion of \(\order\brak{c^{-3}} \) torsion, we do not require any off-diagonal metric components. For our purposes, the connection may be taken to be
\begin{subequations}
\begin{align}
  \Gamma^{i}_{00}
  &=
  c^{-2}\partial^{i}V
  +
  4\pi G c^{-3} \mu^{i}
  \\
  \Gamma^{0}_{0i}
  &=
  c^{-2}\partial_{i}V
  =
  \Gamma^{0}_{i0}
  \\
  \Gamma^{i}_{0j}
  &=
  -4\pi G c^{-3} \theta^{i}{}_{j}
  =
  \Gamma^{i}_{j0}
  \\
  \Gamma^{k}_{ij}
  &=
  4\pi G c^{-3} (\mu_{i}\delta_{j}{}^{k}-\mu^{k}\delta_{ij}) ,
\end{align}
\end{subequations}
with \( \Gamma^{\sigma}_{j0} = \Gamma^{\sigma}_{0j} \).
}

\textcolor{black}{  
Notice how differently \( p^{i} \) and \( q^{i} \) behave in eq.~\eqref{eq:OnePointFiveFinal}. In eq.~\eqref{eq:OnePointFiveFinal__Mom} the former combines with \( \rho\,v^{i} \) to form the `spin-corrected 3-momentum', whilst in eq.~\eqref{eq:OnePointFiveFinal__Mass} \( \rho\,v^{i} \) combines with the \textit{latter} to give the `spin-corrected mass flux' that sets the time-derivative of \( \rho \). Moreover, from eq.~\eqref{eq:OnePointFiveFinal__Mom} we see that the fact that (in general) \( p \neq q \) only contributes further 
to the asymmetry of the 3-stress.
We also remark that these equations introduce further mutual algebraic coupling between \( \mu^{i} \) and \( \theta^{ij} \), with equations \eqref{eq:OnePointFiveFinal__theta} and \eqref{eq:OnePointFiveFinal__mu} \wording{resembling each other closely}.
}

\textcolor{black}{
Importantly, in working to 1.5PN order we have not only reproduced Cosserat elasticity at lowest-order, we have also shown that torsion produces modifications to Cosserat theory when the continuum \wording{becomes weakly relativistic}. 
In particular, we have derived a gravitational torque
\begin{align}
  -(\mu_{i\,}\partial_{j}V-\mu_{j\,}\partial_{i}V)/c
\end{align}
in eq.~\eqref{eq:OnePointFiveFinal__theta}, as well as augmenting eq.~\eqref{eq:OnePointFiveFinal__Mom}'s gravitational force by
\begin{align}
  -
  \mu_{j}\,\partial_{j}\partial_{i}V/c \quad .
\end{align}
These new terms, especially the force, constitute hypotheses that could in principle be tested experimentally. 
At present such tests would presumably be very hard, but if they could ever be performed then they might provide a useful counterpoint to experiments based on gravitational waves.
}


\section{Discussion}
\label{section:discussion}


We have shown that it is possible to reduce the equations of {\ECGR} to those of classical continuum mechanics with spin, or `Cosserat elasticity'. We achieved this by assuming an empirically reasonable scaling for the non-classical degrees of freedom \( \mu^{i} \), \( p^{i} \), \( q^{i} \) and \( \massMom^{ij} \), and then considering the weak-field limit of {\ECGR}. This procedure was helped by the fact that weak-field {\ECGR} reduces to weak-field {\EGR} but with an extra equation representing spin conservation. In addition to the equations of Cosserat theory, our linearisation procedure produced a \wording{classical-looking} equation for the evolution of the mass-moment density.

We also argued that the 1.5PN equations could be derived without much extra effort (as long as one has computer algebra software handy!) and saw from those equations how spin and energy-momentum start to couple together at low PN order. Those equations also drew a clear distinction between the external force and torque due to gravity: a gravitational \textit{force} appears in the equation for linear-momentum conservation at 1PN order, while such a \textit{torque} only enters the equation for spin conservation at 1.5PN.

\textcolor{black}{
The 1.5PN equations contain spin-related terms that are not present in the equations of `pure' Cosserat elasticity. If extremely precise measurements of such a medium's motion were carried out, then in principle one could test for the presence of torsion.
}

The chosen \( c \)-scalings for the non-classical fields are in some sense a postulate of this paper. One may derive weak-field equations of motion where all the non-classical fields are `bumped up' by one power of \( c \), so that 
(for slowly moving matter) 
\( T^{0j} = c(\rho\,v^{j} + q^{j}) \) 
for instance. We did this at first but found that those equations mixed classical and non-classical fields in a way that seemed unphysical. We certainly did not find the equations of Cosserat elasticity. Therefore, we proceeded on the assumption that non-classical fields should not enter the continuum theory.

Incidentally, it is not clear that one may \wording{consistently} import standard linearised {\EGR} techniques using those different scalings. This is because the spin-conservation equation would involve the metric via the Christoffel symbols, so it is not obvious (at least to us) that the linearised {\ECGR} would split into a problem of linearised {\EGR} and an extra equation. Hence, one would probably need to work harder to gauge-fix properly \citep{Barrientos2019}. Perhaps a proper gauge-fixing in this manner would lead to the equations of Cosserat theory even with the different scalings? If it does, we have not worked out how. Hence, we postulate that the non-classical fields should scale with \( c \) as they do.

\begin{acknowledgments}
MM's work on this paper has been supported by the European Research Council (agreement 833848-UEMHP) under the Horizon 2020 programme, and more recently by SNF grant 10.000.683. MM also thanks Marco Loncar for useful conversations. JT gratefully acknowledges input and advice from James Stone. This manuscript has been published in Physical Review D under https://doi.org/10.1103/PhysRevD.109.104052.
\end{acknowledgments}

%
%
%
%

%

\bibliography{CM}

\pagebreak
\appendix

\section{Solution of the weak-field equations}
\label{sec:weakSolution}

Here we solve eqs.~(\ref{eq:effectiveEFE_weak},~\ref{eq:effectiveConservation_weak}). First we reduce the modified EFEs~\eqref{eq:effectiveEFE_weak} to a suggestive weak-field form, following the approach of \citet[][\S 5.5.4]{poisson2014gravity} (although our notation is that of \citet[][\S 7.2]{Carroll}).
Then we choose a convenient gauge and solve the field equations for the necessary components of the metric and connection. Finally, we use those results to write down the weak-field equations.

\subsection{Decomposing the stress-energy field equations}
\label{sec:decomp}

Let us write \( h_{\mu\nu} \) using the irreducible decomposition
\begin{equation}
    h_{00}=\mbox{}-2\,\Phi 
    \quad ,
    \label{eq:c1}
\end{equation}
\begin{equation}
    h_{0i}=w_i
    \quad ,
    \label{eq:c2}
\end{equation}
\begin{equation}
    h_{ij}=2\,s_{ij}-2\,\Psi\,\delta_{ij}
    \quad ,
    \label{eq:c3}
\end{equation}
where we have defined
\begin{equation}
    \Psi=\mbox{}-\tfrac{1}{6}\,\delta^{ij}\,h_{ij}
    \quad ,
    \label{eq:c4}
\end{equation}
and the spatially traceless tensor
\begin{equation}
    s_{ij}=\tfrac{1}{2}\,(h_{ij}-\tfrac{1}{3}\,h\,\delta_{ij})
    \quad .
    \label{eq:c5}
\end{equation}
This decomposition leads to the following representation of the metric part of the Einstein tensor (Carroll's eq.~7.29):
\begin{equation}
\hat{G}{}_{00} =
\partial_i\partial_js^{ij}
+2\,\nabla^2\Psi
\quad ,
\label{eq:Gapproxh002}
\end{equation}
\begin{equation}
\hat{G}{}_{0i} =
\mbox{}-\tfrac{1}{2}\,\nabla^2 w_i
+\tfrac{1}{2} \,\partial_{i} \partial_j w^j
+\partial_{0} \partial_js_{i}{}^{j}
+2\,\partial_{0} \partial_i\Psi
\quad ,
\label{eq:Gapproxh0i2}
\end{equation}
\begin{equation}
\begin{split}
\hat{G}{}_{ij} = & \,  
-\tfrac{1}{2} \,\partial_0(\partial_{i}  w_j
+\partial_{j} w_i
-2\,\delta_{ij}\,\partial_k w^k)
\\ & \mbox{} 
+\partial_{i} \partial_k s_{j}{}^{k}
+\partial_{j} \partial_k s_{i}{}^{k}
-\delta_{ij}\,\partial_k\partial_m s^{km}
-\square s_{ij}
\\ & \mbox{} 
+2\,\partial_0^2 \Psi\,\delta_{ij}
-(\partial_i\partial_{j}-\delta_{ij}\,\nabla^2) (\Phi-\Psi)
\quad .
\end{split}
\label{eq:Gapproxhij2}
\end{equation}

As \cite{Carroll,poisson2014gravity} subsequently discuss, one may further decompose the vector-field \( w_{i} \) and the trace-free tensor-field \( s_{ij} \). We write \( w_{i} \) as
\begin{equation}
    w^i = \partial_{i}\lambda + w_\perp^i
    \label{eq:wdecomp}
\quad ,
\end{equation}
where the \textit{transverse} part \( w_{\perp}^{i} \) has vanishing three-divergence. Similarly, we have the decomposition
\begin{equation}
    s^{ij}=s_\perp^{ij}+s_S^{ij}+s_{\|}^{ij}
    \quad ,
    \label{eq:sijdecomp}
\end{equation}
where~$s_\perp^{ij}$ is a divergence-free transverse part,
\begin{equation}
    \partial_js_\perp^{ij}=0
    \quad ,
\end{equation}
$s^S_{ij}$ denotes the solenoidal part
\begin{equation}
    s^S_{ij}=\tfrac{1}{2}\,(\partial_i\zeta_j+\partial_j\zeta_i)
    \quad ,
        \qquad 
    \partial_i\zeta^i=0
    \quad ,
    \label{eq:Ssolenoidal}
\end{equation}
and~$s^{\|}_{ij}$ denotes the longitudinal part
\begin{equation}
    s^{\|}_{ij}=\left(\partial_i\partial_j-\tfrac{1}{3}\,\delta_{ij}\,\nabla^2\right)\theta
    \quad .
\end{equation}
It is useful to define
\begin{align}
    \tilde{\Psi}
    &=
    \Psi +\tfrac{1}{3}\,\nabla^2\theta
    \quad ,
    \label{eq:Psitilde}
    \\
    \tilde{\Phi}
    &=
    \Phi
    +\partial_{0}\lambda
    -\partial_0^2\theta
    \quad ,
    \label{eq:Phitilde}
    \\
    \wtp 
    &=
    \wp - \partial_{0}\zeta \quad,
\end{align}
in terms of which \( \Ghat_{\mu\nu} \) will take a particularly simple form.

Using these decompositions, the effective EFE~\eqref{eq:effectiveEFE} can be arranged as
\begin{subequations}
\label{eq:EFE_decomp_interm}
\begin{align}
  &
  2\,\nabla^2 \tilde{\Psi}
  =
  \kappa\,\Theta{}_{00} \quad,
  \label{eq:EFE_decomp_00_interm}
  \\
  &
  2\,\nabla^2(\tilde{\Phi} -\tilde{\Psi})
  +
  6\,\partial_0^2 \tilde{\Psi}
  =
  \kappa\,\Theta{}_i{}^i\quad,
  \\
  &
  2\,\partial_0\partial_{i}\tilde{\Psi}
  -\tfrac{1}{2}\,\nabla^2 \wtp_{i}
  =
  \kappa\,\Theta{}_{0i} \quad.
  \label{eq:EFE_decomp_0i_interm}
\end{align}
\end{subequations}
The full spatial part
\begin{align}
  &
  2\,\delta_{ij}\,\partial_{0}^{2}\tilde{\Psi}
  -\tfrac{1}{2}\,
  \partial_{0}(
    \partial_{i}\wtp_{j}
    +
    \partial_{j}\wtp_{i}
  )
  \nonumber\\
  & 
  \mbox{} 
  -\left(\partial_i\partial_j-\delta_{ij}\,\nabla^2\right)
   (\tilde{\Phi} -\tilde{\Psi})
  -\square s_\perp^{ij}
  =
  \kappa\,\Theta{}_{ij} 
\end{align}
describes gravitational radiation, and is therefore unnecessary for our purposes.

Finally, we note that eq.~\eqref{eq:EFE_decomp_0i_interm}'s LHS constitutes a Helmholtz decomposition into the curl-free term \( 2\,\partial_0\partial_{i}\tilde{\Psi} \) and the divergence-free term  \( -\tfrac{1}{2}\,\nabla^2 \wtp_{i} \). Moreover, the timelike component of \wording{the linearised conservation equation} is
\( \partial_{0}\Theta_{00} = \partial_{i}\Theta_{0i} \)\,, wherein we have neglected Christoffel terms at leading order;
this may be substituted into the divergence of eq.~\eqref{eq:EFE_decomp_0i_interm} to give
\begin{align}
  2\,\nabla^2 \partial_{0}\tilde{\Psi}
  =
  \kappa\,\partial_{0}\Theta{}_{00} \quad, 
\end{align}
which holds trivially due to eq.~\eqref{eq:EFE_decomp_00_interm}. We therefore drop eq.~\eqref{eq:EFE_decomp_0i_interm}'s curl-free part, and define \( \Thp_{0i} \) to be \( \Theta_{0i} \)'s divergence-free part. This leaves \wording{our minimal subset decomposed EFEs} as
\begin{subequations}
\label{eq:EFE_decomp}
\begin{align}
  &
  \nabla^2 \tilde{\Psi}
  =
  \tfrac{1}{2}\,\kappa\,\Theta{}_{00} \quad,
  \label{eq:EFE_decomp_00}
  \\
  &
  \nabla^2(\tilde{\Phi} -\tilde{\Psi})
  =
  \tfrac{1}{2}\kappa\,\Theta{}_i{}^i
  -
  3\,\partial_0^2 \tilde{\Psi}\quad,
  \\
  &
  \nabla^2 \wtp_{i}
  =
  \mbox{}-2\,\kappa\,\Thp_{0i} \quad,
  \label{eq:EFE_decomp_0i}
\end{align}
\end{subequations}
with \( \Thp_{0i} \) the transverse part of \( \Theta_{0i} \)\,.

\subsection{A convenient choice of gauge}
\label{section:gauge}

For our present purposes, it is convenient to choose a gauge within which \( \nabla^{2}\Phi = 4 \pi G\rho/c^{2} \). To that end, we impose the following conditions:
\begin{subequations}
\label{eqs:gaugeChoices}
\begin{align}
  &
  \nabla^{4}\theta 
  =
  0 \quad,
  \label{eqs:gaugeChoicesTheta}  
  \\
  &
  \nabla^{2}\partial_{0}\lambda
  =
  \tfrac{1}{2}\,\kappa\,\Theta{}^{i}_{i}
  -
  3\,\partial_{0}^{2}\Psi \quad,
  \label{eqs:gaugeChoicesLambda}\\
  &
  \nabla^{2}\partial_{0}\zeta_{i}
  =
  2\,\kappa\,\Thp_{0i} \quad.
  \label{eqs:gaugeChoicesZeta}
\end{align}
\end{subequations}
These lead to the following gauged equations:
\begin{subequations}
\label{eqs:gaugedEquations}
\begin{align}
  &
  \nabla^{2}\Psi 
  = 
  \frac{4 \pi G}{c^2}\,\rho\quad,
  \label{eqs:gaugedEquationsPsi}\\
  &
  \nabla^{2}(\Phi-\Psi) = 0\quad,
  \label{eqs:gaugedEquationsPhi}\\
  &
  \nabla^{2}\wp_{i} = 0 \quad.
  \label{eqs:gaugedEquationsW}
\end{align}
\end{subequations}
Eq.~\eqref{eqs:gaugeChoicesTheta} sets eq.~\eqref{eqs:gaugedEquationsPsi}, as well as causing \( \theta \) to vanish. Then through eq.~\eqref{eqs:gaugeChoicesLambda} we obtain eq.~\eqref{eqs:gaugedEquationsPhi}, thus enforcing strict equality of \( \Phi \) and \( \Psi \). 
Eq.~\eqref{eqs:gaugeChoicesZeta} is chosen because it leads to eq.~\eqref{eqs:gaugedEquationsW} causing \( \wp_{i} \) to vanish.

\subsection{Solution for metric and connection} 
\label{section:weakmetric}

\( \Phi \) is the only component of the metric itself that we will need within Section~\ref{section:eqmon}. Also, the only component of the connection required within Section~\ref{section:eqmon} is the Christoffel symbol \( \hat{\Gamma}^{i}_{00} \), which is given by
\begin{align}
  \hat{\Gamma}^{i}_{00}
  =
  \partial_{i}\Phi 
  +
  \partial_{0}w^{\perp}_{i}
  +
  \partial_{i}\partial_{0}\lambda \quad,
\end{align}
to leading 
order.
Given that \( \wp_{i} \) vanishes, we must now find \( \Phi \) and \( \lambda \).

To find \( \Phi \), we recall that eq.~\eqref{eqs:gaugedEquationsPhi} sets \( \Psi = \Phi \), so 
\begin{align}
\label{eq:poissonEqFinal}
  \nabla^{2}\Phi = \frac{4 \pi G}{c^2}\rho \quad.
\end{align}
This shows that \( \Phi \) 
has PN order
\begin{align}
  \Phi &\sim \order(c^{-2})\quad.
\end{align}
Next, to deal with \( \lambda \) we note that the order of \( \Phi \) and \( ?\Theta_{i}^{i}? \) lead to eq.~\eqref{eqs:gaugeChoicesLambda} having a RHS of order \( c^{-4} \). It follows that
\begin{align}
  \partial_{0}\lambda &\sim \order(c^{-4}) \quad.
\end{align}
As stated earlier, we only require \( \hat{\Gamma}^{i}_{00} \) accurate to \( \order(c^{-2}) \), so we may neglect \( \lambda \). Since \( \wp_{i} \) vanishes, we conclude that
\begin{align}
\label{eq:Gamma_i00}
  \hat{\Gamma}^{i}_{00} = \partial_{i}\Phi \quad.
\end{align}

\subsection{Conservation of \( \Theta_{\mu\nu} \)} 
\label{section:weakConservationApp}

To obtain the 1-PN equation of conservation of \( \Theta_{\mu\nu} \)\,, we rewrite 
eq.~(\ref{eq:spinConservationPN}) 
explicitly as
\begin{align}
  &
  \partial_{0}\Theta^{\mu 0}
  +
  \partial_{k}\Theta^{\mu k}
  +
  \hat{\Gamma}^{\mu}_{\nu\rho}\Theta^{\rho\nu}
  +
  \hat{\Gamma}^{\nu}_{\nu\rho}\Theta^{\mu\rho}
  =
  0 \quad .
  \label{eq:weakFieldDynamicThetaApp}
\end{align}
Recall that \( \Theta_{\mu\nu} \sim \order(c^{2}) \) and \( \hat{\Gamma}^{\sigma}_{\mu\nu} \sim \order(c^{-2}) \); this means that the \( \Gamma\Theta \) cross-terms in the energy--momentum equation are at most \( \order(c^{0}) \).
Eq.~\eqref{eq:weakFieldDynamicThetaApp}'s timelike component's leading order terms are at \( \order(c) \). The connection therefore plays no role and we find
\begin{align}
\label{eq:massConsWeakApp}
  \partial_{t}\rho
  +
  \partial_{j}(\rho\,v^{j})
  = 0 \quad.
\end{align}
As for eq.~\eqref{eq:weakFieldDynamicThetaApp}'s spacelike components, the leading balance is at \( \order(c^{0}) \), so the only \( \Gamma\Theta \) term that can contribute is \( \hat{\Gamma}^{i}_{00}\Theta^{00} \). From eq.~\eqref{eq:Gamma_i00} this leads to
\begin{align}
\label{eq:threeMomentumConsWeakApp}
  \partial_{t}(
    \rho\,v^{i} 
    - 
    \tfrac{1}{2}\,\partial_{k}\theta^{ki}
  )
  +
  \partial_{j}(
    \partial_{k}\CoupStress^{k(ij)}
    -
    \Sigma^{(ij)}
  )
  =
  \mbox{}-\rho\,\partial^{i}V \quad,
\end{align}
where we have defined
\begin{align}
  V = \Phi \,c^{2} \quad .
\end{align}
Equations~\eqref{eq:massConsWeakApp} and \eqref{eq:threeMomentumConsWeakApp} are as adapted from \cite{poisson2014gravity} in the main text.

\begin{widetext}

\section{1.5PN}
\label{sec:OnePointFiveAlgebra}

Expanding eq.~\eqref{eq:ThetaDefn}, we find after tedious, computer-assisted algebra that
\begin{align}
  \Theta^{00}
  &=
  c^{2}\rho 
  -
  c\partial_{i}\mu^{i}
  +
  \order(c^{0})
  \\
  \Theta^{0j}
  &=
  c(
    \rho\,v^{j} 
    + 
    \tfrac{1}{2}\,\partial_{k}\theta^{jk}
  )
  +
  \tfrac{1}{2}\,c^{0}[
    (p^{j}+q^{j})
    +
    \partial_{k}\MassMom^{kj}
    +
    \partial_t{\mu}^{j}
    -
    2\,\partial_{k}\mu^{[k}v^{j]}
  ]
  +
  \order(c^{-1})
  \\
  \Theta^{ij}
  &=
  (
    \partial_{k}\CoupStress^{k(ij)} 
    - 
    \Sigma^{ij}
  )
  +
  c^{-1}[
    (p+q)^{(i}v^{j)}
    -
    \partial_t{\MassMom}^{(ij)}
    +
    \partial_{k}
    (
      \MassMom^{k(i}v^{j)}
      -
      v^{k}\MassMom^{(ij)}  
    )
    +
    \christMTheta^{ij}
  ]
  +
  \order(c^{-2}) \quad,
  \label{eq:OnePointFiveTheta_ij_init}
\end{align}
with
\begin{align}
  \christMTheta_{ij}
  =\mbox{}
  \mu_{(i\,}\partial_{j)}V \quad.
\end{align}
Now we expand eq.~\eqref{eq:spinConservation} to give
\begin{align}
  &
  \mbox{}-\partial_{t}(\mu^{i}+\theta^{ij}v_{j}/c)
  +\partial_{j}(\MassMom^{ij}-\CoupStress^{kij}v_{k}/c)
  +\christmu^{i}/c
  =
  q^{i}-p^{i}
  +
  2\,\Sigma^{[ij]}v_{j}/c
  +
  \order(c^{-2})
  \label{eq:muOnePointFive}
  \\
  &
  +\partial_{t}(\theta^{ij} + 2\,\mu^{[i}v^{j]}/c)
  -\partial_{k}(\CoupStress^{ijk}-2\,v^{[i}\MassMom^{j]k}/c)
  +\christmu^{ij}/c
  =
  -2\,\Sigma^{[ij]}
  +
  2\,(p-q)^{[i}v^{j]}/c
  +
  \order(c^{-2}) \quad,
  \label{eq:thetaOnePointFive}
\end{align}
where the terms
\begin{align}
  \christmu^{i} 
  &= 
  -\theta^{ij}\partial_{j}V \quad ,
  \\
  \christtheta_{ij}
  &=
  -2\,\mu_{[j\,}\partial_{i]}V \quad ,
\end{align}
arise from the terms in eq.~\eqref{eq:spinConservation} involving (metric) Christoffel symbols.

We need to compute various derivatives of \( \Theta \) to expand eq.~\eqref{eq:effectiveConservation}. We go sequentially, making judicious use of the evolution equations for \( \mu^{i} \) and \( \theta^{ij} \). First, we note that
\begin{align}
  \partial_t{\Theta}^{00}
  =
  c^{2}\partial_t\rho
  -
  c\,\partial_{i}(
    \partial_{j}\MassMom^{ij}
    +
    p^{i}
    -
    q^{i}    
  )
  +
  \order(c^{0}) \quad,
\end{align}
on account of eq.~\eqref{eq:muOnePointFive}. Eq.~\eqref{eq:muOnePointFive} also implies that
\begin{align}
\label{eq:ThetaZeroJOnePointFiveSimp}
  \Theta^{0j}
  =
  c(\rho\,v^{j} + \tfrac{1}{2}\partial_{k}\theta^{jk})
  +
  c^{0}[
    p^{j}
    +
    \partial_{k}\MassMom^{(jk)}
    -
    \partial_{k}(\mu^{[k}v^{j]})
  ]
  +
  \order(c^{-1}) \quad.
\end{align}

We now have enough to state conservation of mass. Taking account of various terms' (anti)symmetry, we find that eq.~\eqref{eq:effectiveConservation}'s zero-component is
\begin{align}
  \covmet{\nu}\Theta^{0\nu}
  &=
  c^{-1}\partial_t{\Theta}^{00}
  +
  \partial_{j}\Theta^{0j}
    +
  \order(c^{-2})
  \nonumber\\
  &=
  \partial_t\rho
  +
  \partial_{j}(
    \rho\,v^{j} +q^{j}/c
  )
  +
  \order(c^{-2})
  = 
  0 \quad.
  \label{eq:rhoOnePointFive}
\end{align}
At this order, the Christoffel symbols do not contribute.

To write down conservation of momentum, we must manipulate \( \partial_{j}\Theta^{ij} \) suitably. To that end, we note that
\begin{align}
  \partial_{j}\partial_{k}\CoupStress^{k(ij)}
  =
  \mbox{}-\tfrac{1}{2}\,\partial_{j}\partial_{k}\CoupStress^{ijk} \quad,
\end{align}
due to \( C \)'s (anti)symmetries. This allows us to substitute eq.~\eqref{eq:thetaOnePointFive} into the three-divergence of eq.~\eqref{eq:OnePointFiveTheta_ij_init}, which after some tedious algebra gives
\begin{align}
  \partial_{j}\Theta^{ij}
  =
  \mbox{}-\tfrac{1}{2}\,\partial_{j}\partial_t\theta^{ij}
  -\partial_{j}\Sigma^{ij}
  +
  c^{-1}\partial_{j}[
    p^{i}v^{j}
    +
    q^{j}v^{i}
    +
    (\christMTheta^{ij}-\tfrac{1}{2}\christtheta^{ij})
    -
    \partial_{t}(v^{[j}\mu^{i]})
    -
    \partial_t{\MassMom}^{(ij)}
  ]
  +
  \order(c^{-2}) \quad.
\end{align}
In getting here it is crucial that the \( v\MassMom \) cross-terms cancel; this happens because
\begin{align}
  \partial_{j}\partial_{k}
  (
    \MassMom^{k(i}v^{j)}
    -
    v^{k}\MassMom^{ij}
    -
    v^{[i}\MassMom^{j]k}
  )
  =
  \partial_{j}\partial_{k}
 (
    v^{[j}\MassMom^{k]i}
    -
    \MassMom^{i[j}v^{k]}
    -
    v^{i}\MassMom^{[jk]}
  )
  = 0 \quad .
\end{align}
Using eq.~\eqref{eq:ThetaZeroJOnePointFiveSimp} we find, after many cancellations, that
\begin{align}
  \partial_{t}(\rho\,v^{i} + p^{i}/c)
  -
  \partial_{j}[
    \Sigma^{ij}
    -
    (
      p^{i}v^{j}
      + 
      q^{j}v^{i}
    )/c
  ]
  +
  \partial_{j}
  (
    \christMTheta^{ij}
    -
    \tfrac{1}{2}\christtheta^{ij}
  )/c
  +\christMom^{i}
  = 
  \order(c^{-2}) \quad.
\end{align}
Here the Christoffel terms from the EFE do contribute:
\begin{align}
  \christMom_{i}
  =
  (
    \rho 
    - 
    \partial_{k}\mu^{k}/c
  )\,
  \partial_{i}V \quad.
\end{align}
It also happens that
\begin{align}
  (
    \christMTheta^{ij}
    -
    \tfrac{1}{2}\christtheta^{ij}
  )
  =
  \mu^{j}\,\partial^{i}V \quad,
\end{align}
whereupon we find that
\begin{align}
  \partial_{t}(\rho\,v_{i} + p_{i}/c)
  -
  \partial_{j}[
    ?\Sigma_{i}^{j}?
    -
    (
      p_{i}v^{j}
      + 
      v_{i}q^{j}
    )/c
  ]
  =
  -\rho\,\partial_{i}V
  -
  \mu^{j}\partial_{j}\partial_{i}V/c
  + 
  \order(c^{-2}) \quad.
  \label{eq:momOnePointFive}
\end{align}
This is the 1.5PN statement of 3-momentum conservation.
  
\end{widetext}

\end{document}